# Magneto-Mechanical Bilayer Metasurface with Global Area-Preserving Density Tunability for Acoustic Wave Regulation

*Jay Sim, Shuai Wu, Jize Dai and Ruike Renee Zhao\**

J. Sim, S. Wu, J. Dai, R. R. Zhao
Department of Mechanical Engineering
Stanford University
Stanford, CA 94305, USA
E-mail: rrzhao@stanford.edu

Metasurfaces have extensive potential in acoustic cloaking, optical scattering, and electromagnetic antenna due to their unprecedented properties and the ability to conform to curved substrates. Active metasurfaces have attracted significant research attention because of their on-demand tunable properties and performances through shape reconfigurations. They normally achieve active properties through internal structural deformations, which often lead to changes in overall dimensions. This also demands the corresponding alterations of the conforming substrate, which could be a significant limitation for their practical applications. To date, achieving area-preserving active metasurfaces with distinct shape reconfigurations remains a prominent challenge. In this paper, we present magneto-mechanical bilayer metasurfaces that demonstrate area density tunability with area-preserving capability. The bilayer metasurfaces primarily consist of two arrays of magnetic soft materials with distinct magnetization distributions. Under an external magnetic field, each layer behaves differently, which allows the metasurface to reconfigure its shape into multiple modes and thus significantly tune its area density without changing its overall dimensions. The area-preserving multimodal shape reconfigurations are further exploited as active acoustic wave regulators to tune bandgaps and wave propagations. The bilayer approach thus provides a new concept to design area-preserving active metasurfaces for broader practical applications.

# 1. Introduction

Metamaterials are manmade materials with a periodic structure comprised of unit cells and are engineered to exhibit extraordinary properties derived from their geometry rather than the material composition. Metasurfaces, or two dimensional (2D) metamaterials, consist of repeating periodic 2D unit cells and are commonly designed to conform to flat or curved substrates for various applications. Metasurface in optics can be designed to scatter[1], modulate the intensity of[2,3], and refract[4] light, creating new methods in light manipulation. Similarly, electromagnetic waves can be scattered into specific shapes[5,6], cloaked[6], steered to form antennae[7,8], or utilized as conformal filters[9]. Acoustic metasurfaces can also be engineered to filter[10,11], isolate[12,13], cloak[14–16], regulate wave transmission[17–19], absorb[20,21], or steer acoustic waves[22–24].

In general, the properties of metasurfaces are determined by their geometry[25]. Conventional metasurfaces have fixed unit cell geometries, and accordingly, fixed properties, which limits their applicability. In response, active metasurfaces have emerged, which can transform between preprogrammed shapes when subjected to external stimuli, such as mechanical loads[26–29], heat[30–32], electrical currents[33–35], or magnetic fields[36–38]. These shape transformations often involve internal deformations of the unit cells, which also change the area density as well as the overall dimensions of the metasurface. However, a metasurface is intended to cover a specific area. If its overall dimensions change dramatically during actuation, the metasurface would fail to provide complete area coverage, which could impose significant limitation for practical applications.

In this paper, we present a new metasurface design strategy that can significantly change the area density while maintaining overall area through a novel bilayer concept. Here, the metasurfaces are made of hard-magnetic soft active materials (hm-SAM) which have a soft elastomer matrix embedded with hard magnetic particles. The material can provide untethered, fast, and reversible actuation by a remotely applied magnetic field[39–41]. The square-shaped unit cell of the array is first designed with a preprogrammed magnetization distribution that can induce contraction for area density change under an external magnetic field. When we overlay two arrays in certain arrangements, due to the magnetic attraction, they form an integrated metasurface with strong interfacial adhesion. Under an external magnetic field, only one layer can significantly contract to provide the area density change, while the other maintains the overall dimensions of the metasurface, a process otherwise unobtainable in a single layer setting.

In addition, the separate layer actuation increases the number of unique shape reconfigurations, and correspondingly, acoustic property modes. Based on this concept, we demonstrate bilayer metasurface designs with different assembling strategies, including an offset arrangement, flipping arrangement, and distributed patches. Their corresponding wave regulation performances, bandgap and waveguide, are investigated. Parametric studies are conducted for the flip-arranged bilayer to explore how the bilayer thickness ratio can be programmed to tune acoustic bandgaps. The ability of the bilayer metasurface to greatly change its area density while retaining constant overall dimensions thus provides a new strategy of designing highly tunable area-preserving conformal acoustic regulators for practical applications.

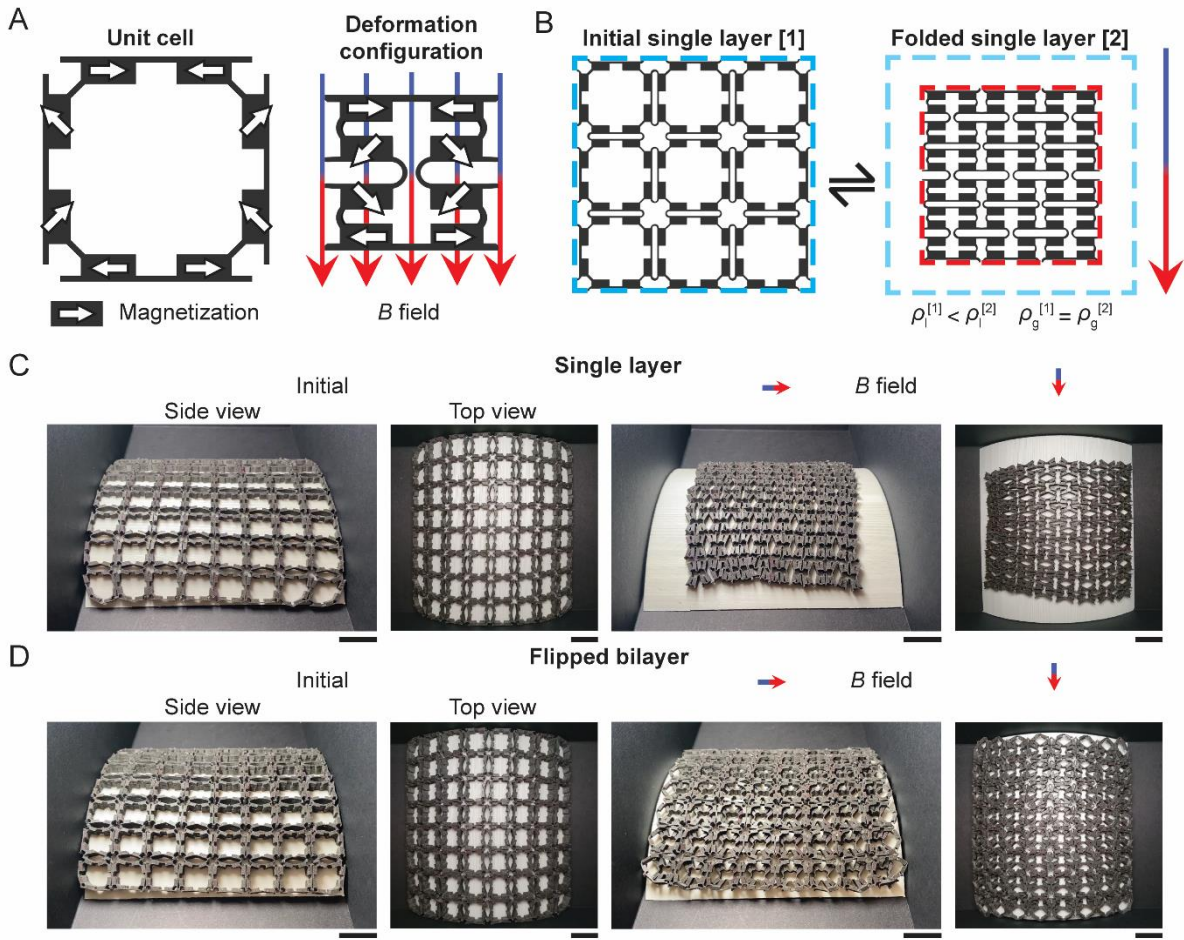

**Figure 1.** Magneto-mechanical single-layer metasurface with area-changing actuation and bilayer metasurface for area-preserving actuation. (A) Magnetization distribution and actuation of a unit cell. (B) Single layer metasurface and its actuation to the folded mode, with their areas denoted by the dashed blue and red box, respectively. The global area density, $\rho_g$, is constant during actuation but the local area density, $\rho_l$, increases. (C) Single-layer metasurface conforms to a curved substrate with incomplete area coverage after actuation. (D) Bilayer metasurface conforms to a curved substrate with complete area coverage throughout actuation. Scale bars: 20 mm.

## 2. Results
### 2.1. Bilayer Metasurface with Multimodal Shape Control

As shown in **Figure 1A**, the square-shaped unit cell design in this work consists of eight rectangular blocks connected to each other with thin hinges. Both hinges and blocks are hm-SAMs and are fabricated using casting methods. Further information on the dimensions and fabrication method can be found in the experimental methods and the Figures S1 and S2 (Supporting Information). The unit cells are magnetized with predetermined magnetization distribution, as denoted by the white arrows, to enable area contraction for density change under an external magnetic field. The upper and lower blocks are magnetized along the block length with alternating directions and the side blocks are magnetized diagonally at 45º. Here, the magnetization directions were carefully designed to provide special actuation. When a downward magnetic field is applied, the block's magnetization will try to align with the magnetic field due to the generated torque, $\tau = \mathbf{M} \times \mathbf{B}$, where $\mathbf{M}$ is the magnetization density of the blocks, and $\mathbf{B}$ is the magentic field. For the upper and lower blocks, their magnetization is initially 90º to the magnetic field. As the block rotates, the torque decreases from its maximum value. For the side blocks, their magnetization is initially 135º to the magnetic field. The corresponding torque is initially smaller than the upper and lower blocks but increases as they rotate by 45º and then starts to decrease. As a result, the unit cell will try to deform into a cross shape first, and then pack into a much smaller rectangular shape. Individual unit cells can be connected to create a single layer array, as illustrated in **Figure 1B**. Under a downward magnetic field, the single layer actuates in the same manner as the individual unit cell and deforms into the folded mode. The total area of the single layer in the initial configuration is marked by the dashed blue box whilst the actuated configuration is marked by the red dashed box (in the rest of the paper, we will use the dashed blue box for the initial configuration and the red dashed box for actuated configuration). We also define two types of area densities, $\rho_g$ and $\rho_l$, as the global area density and the local area density, which are measured as the 2D material's surface area of the metasurface ($A_{meta}$) divided by the area of the initial configuration ($A_{initial}$) and the total area of the deformed mode ($A_{deform}$), respectively. During the actuation of the single layer metasurface, although its $\rho_l$ significantly increases from 25.55% to 44.97%, when the unit cells turns into a compact shape, its $\rho_g$ remains constant as there is no addition to the material surface area. This leads to a region surrounding the folded mode without material coverage. This also implies the metasurface has to reduce its size to increase its local area density. It should also be noted that under an upward magnetic field, the unit cell will expand[42] . However, this expansion is suppressed in the array by the interaction among neighboring cells. Therefore, under an upward field, there is no obvious area change. **Figure 1C** depicts the challenges in developing a area-preserving metasurface. The single-layer magneto-mechanical metasurface conforms to a curved substrate. When there is no magnetic field, the metasurface covers the entirety of the substrate. But when a magnetic field is applied and the metasurface undergoes actuation, it rapidly contracts and a significant area of the substrate is left uncovered. In **Figure 1D**, a bilayer in a flipped arrangement is placed on the same curved substrate. Even when there is an applied magnetic field, the substrate is still completely covered.

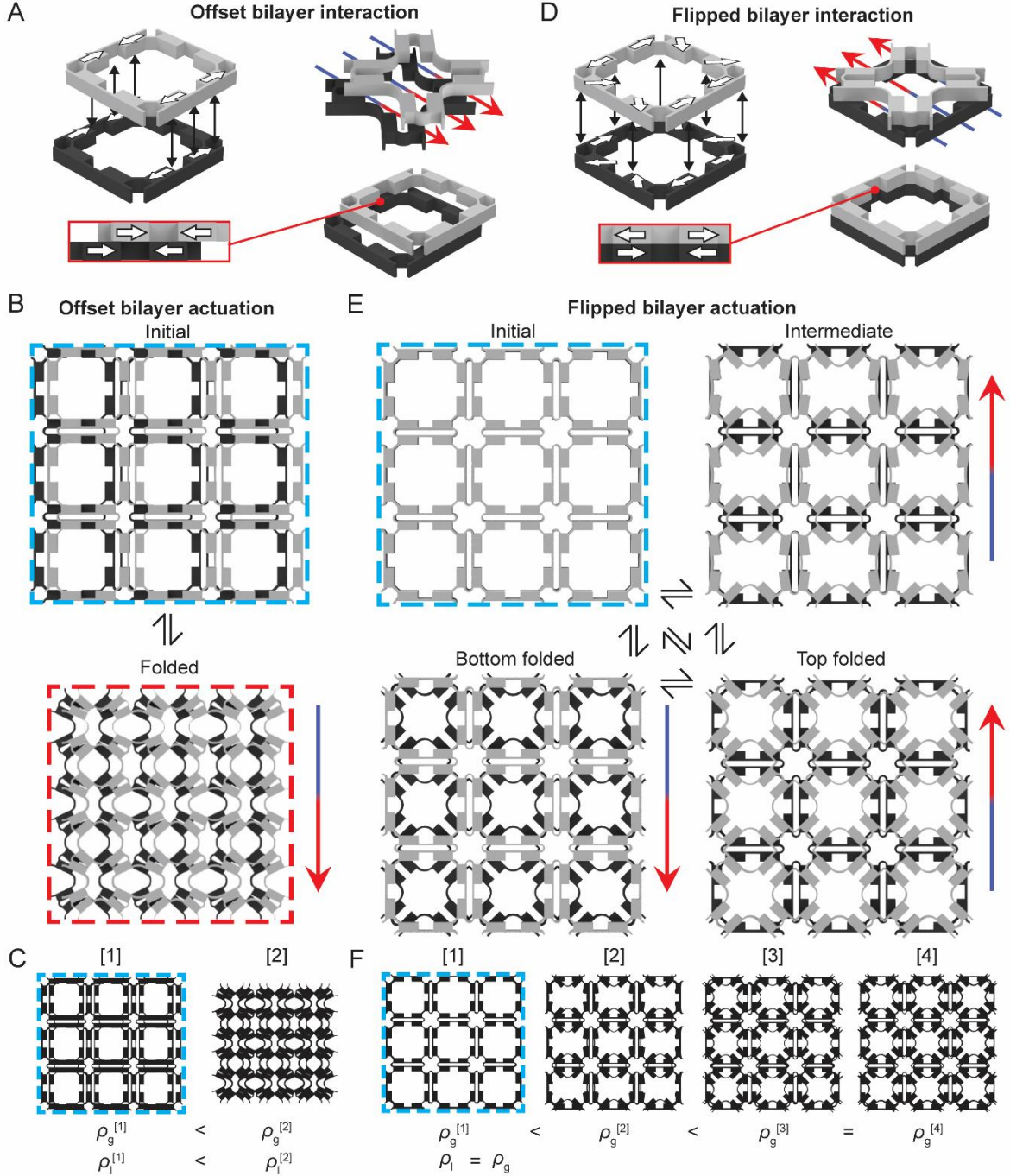

**Figure 2.** Bilayer metasurfaces with two distinct overlay strategies. (A) Magnetic interaction between two unit cells arranged in an offset bilayer and their actuation. (B) A 3 × 3 offset bilayer and its magnetic actuation. The $A_{deform}$ is smaller than $A_{initial}$. (C) Both global and local area densities of the offset bilayer in the folded mode are larger than that of the initial mode. (D) Magnetic interaction between two unit cells arranged in a flipped bilayer and their actuation. (E) A 3 × 3 flipped bilayer assembly and the four magnetically actuated modes, showing area-preserving deformation ($A_{initial} = A_{deform}$). (F) Comparison of the area densites for the four flipped bilayer modes. Because $A_{initial}$ and $A_{deform}$ are the same, their global and local area densities are the same. Area densities increase from the initial mode to the intermediate mode and then the top folded mode. There is no change in area densities between the top folded and bottom folded modes.
6

In this work, we use different ways to overlay two array layers into a bilayer metasurface to effectively tune its area density while preserving its global dimension. In the first approach, guided by the magnetic attraction, the two layers are automatically offset slightly when they are adhered to each other, as shown **Figure 2A**, where the unit cell on top is colored in light gray and the bottom unit cell in black. This metasurface is called the offset bilayer. Under a downward magnetic field, both unit cells would tend to fold into a compact shape because they have the same magnetization distributions. However, the constraint due to the magnetic interaction between the two layers prevents the arrays from becoming fully compacted. As a result, the unit cells transform into a cross shape. **Figure 2B** shows the initial and folded configurations of a $3 \times 3$ offset bilayer metasurface. The folded mode is noticeably smaller than the initial mode, but its dimension change is not as dramatic as the single layer. **Figure 2C** shows the area densities of the offset bilayer in the initial and folded modes. Because of the offset array arrangement, both the global and local area densities will increase during actuation, showing an improving, but not perfect, area-preserving actuation.

The second bilayer arrangement strategy is based on a flipped unit cell placed on top of the original unit without offset in either the horizontal or vertical directions. Most importantly, each block in the top unit cell has a reversed, or flipped magnetization direction of the corresponding block in the bottom one, as presented in **Figure 2D**. This creates a strong magnetic attraction because the individual blocks act as magnetic dipoles stacked in opposite directions. In addition, because the unit cells have flipped magnetization distributions, only one layer can fold, or actuate, and the other one will not change its shape, regardless of the direction of the applied magnetic field. This separately controlled actuation is the key design concept in the flipped bilayer, as shown in **Figure 2E**. In the initial mode, where there is no applied magnetic field, the flipped bilayer metasurface has the same configuration as the single layer. Under an upward magnetic field, only the upper and lower blocks of the top layer fold while the side blocks are largely undeformed. This mode is referred to as the intermediate mode and occurs because of the magnetization design of the unit cell and the attractive interaction between the two layers. As discussed above, the magnetization of upper and lower blocks is aligned in the horizontal direction and therefore, is perpendicular to the applied field and provides the largest torque. In comparison, the angle between the magnetization in the side blocks and the field is 135°, which provides smaller torques. The magnitude of torque is proportional to $\sin\theta$, where $\theta$ is the angle between the magnetization and the applied field. Furthermore, at this mode, the bottom layer remains unfolded. The strong magnetic attraction between the top and bottom array constrains the overall dimension change of the top layer, while allowing its local shape reconfiguration. At an increased upward field, the torque from the side blocks becomes sufficient to overcome the magnetic attraction between top and bottom and undergo shape reconfiguration into a cross shape, which is the folded mode. Similar to previous actuation, there is no obvious overall dimension change of the metasurface. Under a downward magnetic field, the actuations are reversed, where the bottom layer folds and the top layer remains unfolded. Again, the separate actuation of layers provides folding and large shape deformation while preserving the global dimension of the metasurface. In **Figure 2F**, we show the area density change of the flipped bilayer. As discussed above, in the bilayer case, the blue box and the red box have the same area; therefore, the global area density is the same as the local area density. We term the in-plane density of the initial mode as $\rho_1$. At the initial mode, the 2D area of the metasurface is the same as a single layer. For the area density of the intermediate mode, or $\rho_2$, because the top layer partially folds, increasing the metasurface's material area, we have $\rho_2 > \rho_1$. In the top folded mode, because the remaining parts of the top layer fold inwards, increasing the material area further, its area density, $\rho_3$, is larger than $\rho_2$. For the complete bottom folded mode, since it is a reverse of the top one, its area density, $\rho_4$ is the same as $\rho_3$.

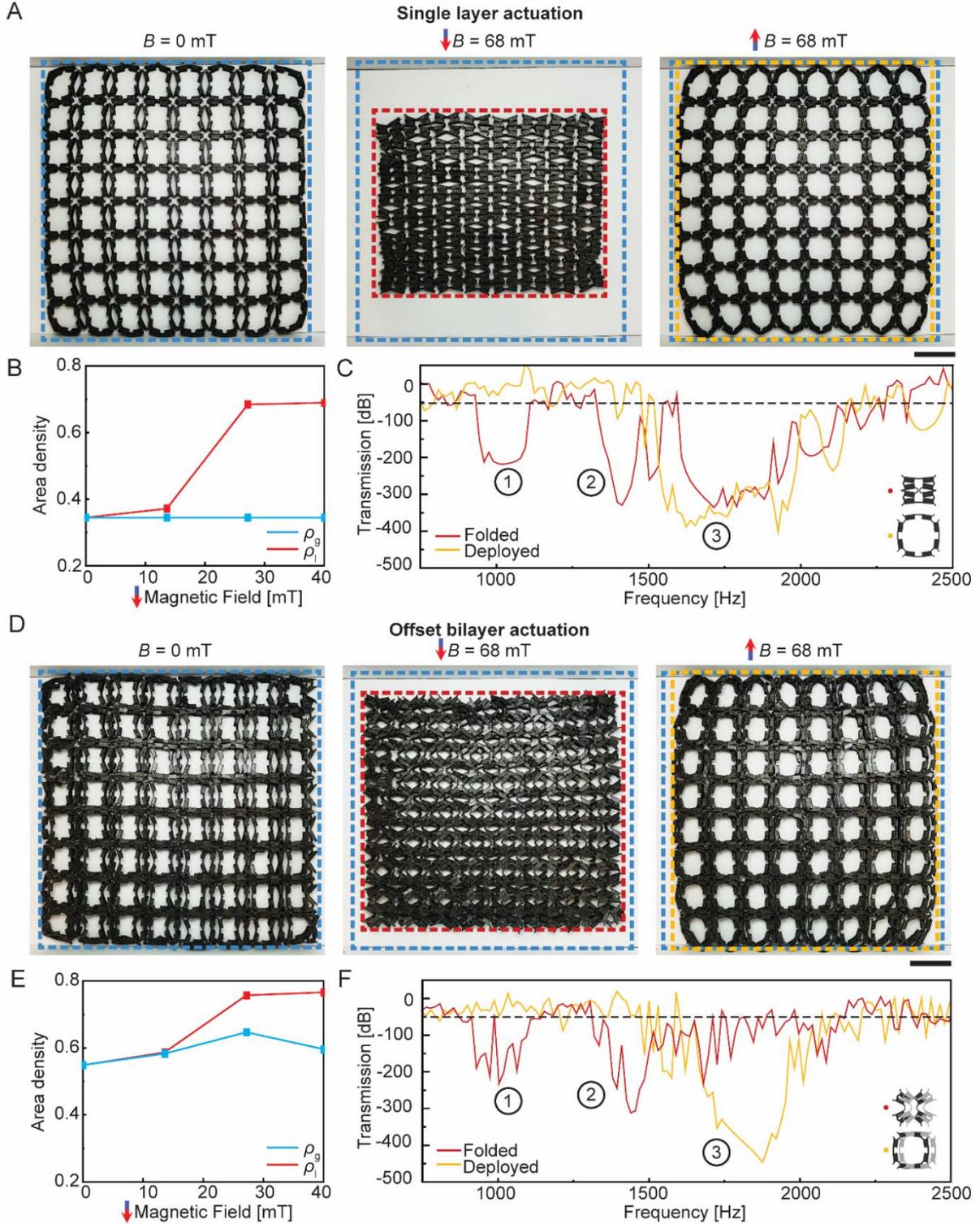

**Figure 3.** Metasurfaces made of a single layer array and two offset layers. (A) Actuation of the single layer magneto-mechanical metasurface. Scale bar: 20 mm. (B) Global and local area densities of the single layer metasurface versus downward magnetic field. (C) Frequency versus transmission curves of the single layer metasurface in the folded and deployed modes. (D) Actuation of the offset bilayer metasurface. Scale bar: 20 mm. (E) Global and local area densities of the single layer metasurface versus downward magnetic field. (F) Frequency versus transmission curves of the offset bilayer metasurface in the folded and deployed modes.

## 2.2. Bilayer with Horizontal Offset

After discussing the concept of the bilayer designs, we present the experimental results. **Figure 3A** shows the single layer metasurface, which is fabricated using 2 mm thick unit cells with 25 vol% NdFeB particles. The unit cells are arranged in an 8 × 8 grid, for a total of 64 unit cells. See experimental methods and the supporting information for further details on the fabrication. The single layer array is placed between a pair of Helmholtz coils, which provides a single-axis homogeneous magnetic field. See Figure S3 for more details on the experimental setup. At the initial mode, the vertical blocks in the single layer exhibit negligible folding due to small repulsive forces between unit cells. A downward 68 mT magnetic field can fold the metasurface into the folded mode with a significant global dimension reduction. An upward 68 mT magnetic field is applied to overcome the repelling magnetic force in the vertical blocks to fully unfold, or deploy, the metasurface. **Figure 3B** examines the area density change with the applied magnetic field. In the initial mode (B = 0mT), the global and local densities are the same and are at 0.344. The local density increases drastically with increasing downward magnetic field, due to the folding mode. However, since there is no material addition or reduction, the global density is constant during actuation. Note that an upward magnetic field does not change the shape nor the area densities drastically (See Figure S4 for more details and experimental methods for the calculation of the area densities).

We further study how the actuation of the single layer metasurface affects its acoustic properties, simulated by finite element analysis (FEA). Further information on acoustic simulation can be found in the supporting information and necessary material characterization for the acoustic simulation can be found in figures S5 and S6. **Figure 3C** shows the transmission of waves at different frequencies. Here, transmission drops of at least 50 dB were considered as preventing waves from propagation. Three frequency regions (bandgaps) are marked in black circles. In the first and second frequency regions, at 929 Hz – 1110 Hz and 1321 Hz – 1562 Hz, the folded metasurface displays significant transmission drops while the metasurface in the deployed mode does not. At the third frequency region, from 1517 Hz – 2150 Hz, both the folded mode and the deployed mode exhibit transmission drops of at least 300 dB.

In **Figure 3D**, the offset bilayer consists of a top layer of 64 unit cells in an 8 × 8 grid. These unit cells are 2 mm thick with 25 vol% NdFeB particles. The bottom layer has the same number of unit cellsas the top, but is 4 mm thick with 30 vol% NdFeB particles. The offset bilayer was also placed between the Helmholtz coils and the same magnetic field as the single layer assembly was applied to achieve three modes: initial, folded, and deployed. When a downward magnetic field of 68 mT is applied, the offset bilayer contracts into the folded mode. As discussed above, because of the magnetic attraction within the bilayer, the unit cells fold into a cross-like shape. **Figure 3E** illustrates the area densities of the offset bilayer. The initial mode exhibits an area density of 0.548, which is larger than the density of the single layer due to the offset arrangment. During folding under a downward magnetic field, the local density increases to 0.766, while the global density also increases to 0.596 at 40 mT. However, the global area density decreases slightly at 27 mT to 40 mT because of the overlaid top and bottom arrays. Still, the global area density increases overall from 0 mT to 40 mT, demonstrating improved area-preserving actuation compared to the single layer metasurface. The complete area densities in both magnetic field directions are shown in figure S7. **Figure 3F** depicts the transmission versus frequency curves of the offset bilayer. Similar to the single layer, three frequency regions are denoted. In the first and second frequency regions, 913Hz - 1153Hz and 1305 Hz to 1740 Hz, only the folded mode show significant transmission drops, which is the same behavior as the single layer in the folded mode. However, unlike the single layer at the third frequency region, the offset bilayer at the folded mode does not show any transmission drops and the deployed mode shows a transmission drop of 447 dB from 1700 Hz – 2000 Hz.

The drastic difference in transmission drops between the folded and deployed modes in the single layer and the offset bilayer can be utilized as a device for switchable acoustic bandgaps. In the single layer, two low frequency bandgaps (frequency regions 1 and 2) can be toggled under a downward magnetic field, and the bandgap that exists at high frequencies (frequency region 3) remains unchanged. For the offset bilayer, two low frequency bandgaps can be switched on but the high frequency bandgap can be toggled off by a downward magnetic field. This acoustic behavior between modes in the offset bilayer is much more dramatic and thus, a more suitable candidate applications that require tunable acoustic properties.

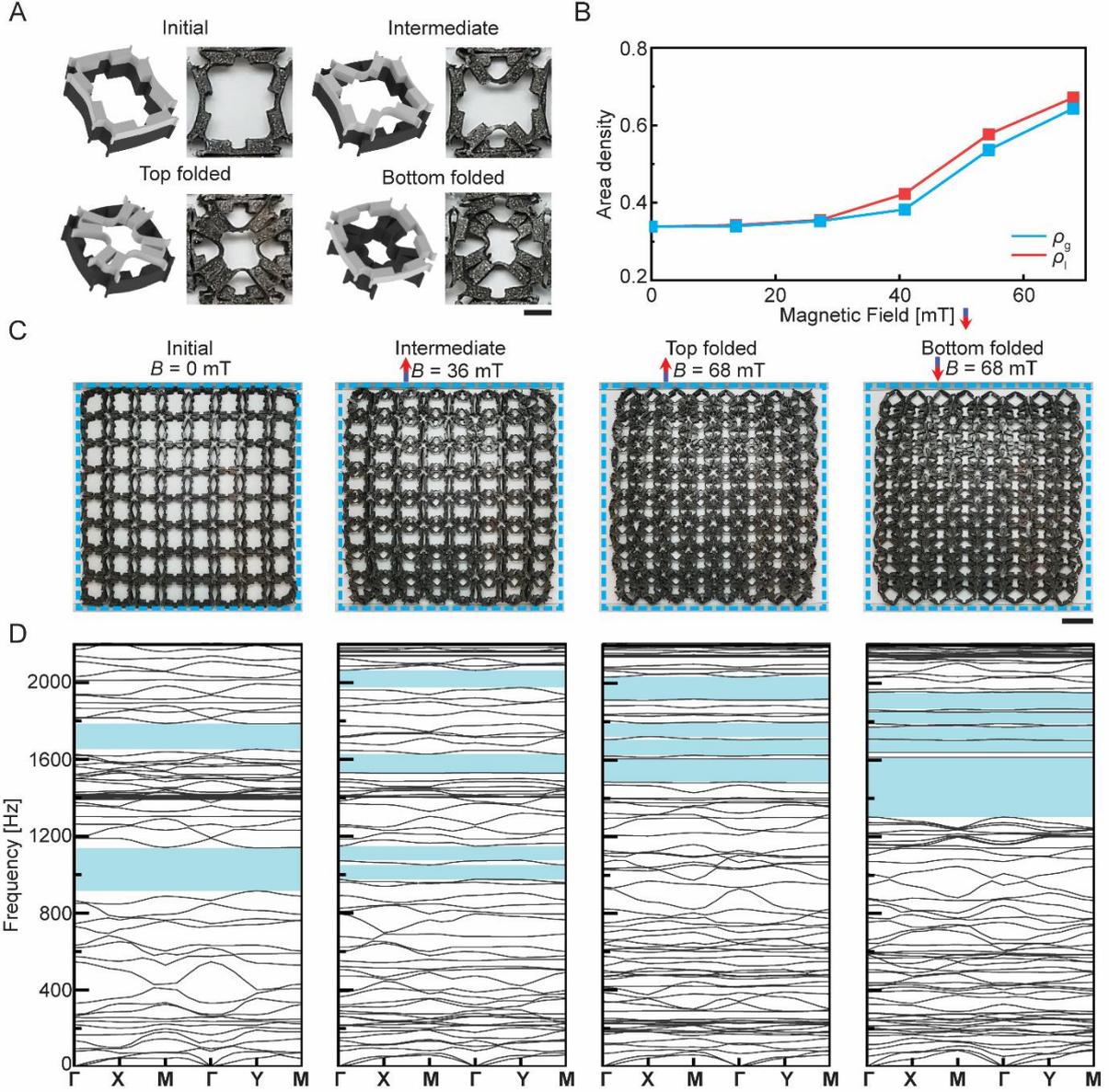

**Figure 4.** Flipped bilayer metasurface. (A) The experimental four modes of the flipped bilayer compared with the 3D model used in the Bloch wave analysis. Scale bar: 5 mm. (B) Global and local area densities versus the downward magnetic field of the flipped bilayer. (C) Experimental actuation of the flipped bilayer. Scale bar: 20 mm. (D) Dispersion curves of modes with highlighted acoustic bandgaps that are at least 50 Hz wide.

## 2.3. Flipped Bilayer

**Figure 4A** shows the schematic and experimental actuation of the unit cells in the four modes in the flipped bilayer metasurface: initial, intermediate, top folded, and bottom folded. The flipped bilayer consists of two layers of 64 unit cells each, arranged in 8 × 8 grids with no offset in horizontal or vertical directions. The top layer is 2 mm thick with 25 vol% NdFeB particles, and the bottom is 4 mm thick with 30 vol% NdFeB particles. The global and local area densities versus the downward magnetic field for the flipped bilayer is illustrated in **Figure 4B**. This bilayer does not have any global dimension change during actuation, so the local area density should be the same as the global area density. At 0 mT, the flipped bilayer has an area density of 0.55. At 68 mT downward, the global and local area densities are 0.63 and 0.65, respectively, showing great agreement. More information on the area densities of the flipped bilayer is shown in Figure S8 The experimental magnetic actuation of the flipped bilayer is shown in **Figure 4C**. An upward magnetic field of 36 mT was applied to actuate the metasurface to the intermediate mode where the upper and lower blocks of the unit cells in the top layer fold inwards. But while the top layer is actuated, the bottom layer does not fold because its unit cells were programmed with reversed magnetization directions. This independent actuation of the top layer allows for distinct shape deformation with negligible displacement, as shown by the unchanged dashed blue box. The field is then increased to 68 mT upward to achieve the complete top folded mode where the blocks in the top layer now fold inward to create a cross shape and the bottom layer remains undeformed. Lastly, the metasurface achieves the bottom folded mode under a downward field of 68 mT. The flipped bilayer displays four distinct modes that can be controlled by adjusting the direction and magnitude of the applied magnetic field. Note that the actuation in the intermediate top mode is not completely uniform and some areas along the edges of the bilayer display complete top folding due to the inhomogeneity of the magnetic field at the edges.

**Figure 4D** shows the shifting acoustic bandgaps, which were calculated by using the Bloch wave analysis method[43] with eigenmode calculations on the representative volume element (RVE). Details on the RVE and the Brillouin zone of the Bloch wave analysis are illustrated in Figure S7. The initial mode of the metasurface exhibits two noticeable bandgaps: 914 Hz – 1138 Hz and 1654 Hz – 1785 Hz. At an upward 36 mT, or the intermediate mode, the first initial mode bandgap splits into two at 976 Hz – 1050 Hz and 1075 Hz – 1143 Hz while the second initial mode bandgap narrows and shifts to 1531 Hz – 1627 Hz. A high frequency bandgap also appears at 1974 Hz – 2065 Hz for a total of four distinct bandgaps. When the magnetic field is increased to 68 mT and the bilayer is in the complete top folded mode, there are four new high frequency bandgaps that appear above 1400 Hz. The first three are at 1486 Hz – 1605 Hz, 1628 Hz – 1706 Hz, and 1723 Hz – 1791 Hz. The last bandgap is at a much higher frequency, 1919 Hz – 2033 Hz. From the intermediate mode to the top folded mode, the low frequency bandgaps disappear and are replaced by two additional high frequency bandgaps. Lastly, when the 68 mT downward magnetic field is applied, the bilayer is in the bottom folded mode and lowest frequency bandgap widens to 1304 Hz – 1616 Hz, the single widest bandgap across all modes. The next two bandgaps narrow slightly to 1641 Hz – 1702 Hz and 1709 Hz – 1772 Hz. Lastly, the high frequency bandgap splits into two and shifts to 1790 Hz – 1851 Hz and 1867 Hz – 1951 Hz for a total of five bandgaps. The acoustic property at the complete bottom folded mode is different from that of the top folded mode because the top and bottom layers have different thicknesses. Overall, the flipped bilayer shows distinct acoustic responses that can be reliably controlled through the magnetic actuation. The unique combination of local area density changes, constant global dimensions, and shifting acoustic response can be used to develop highly tunable acoustic filters.

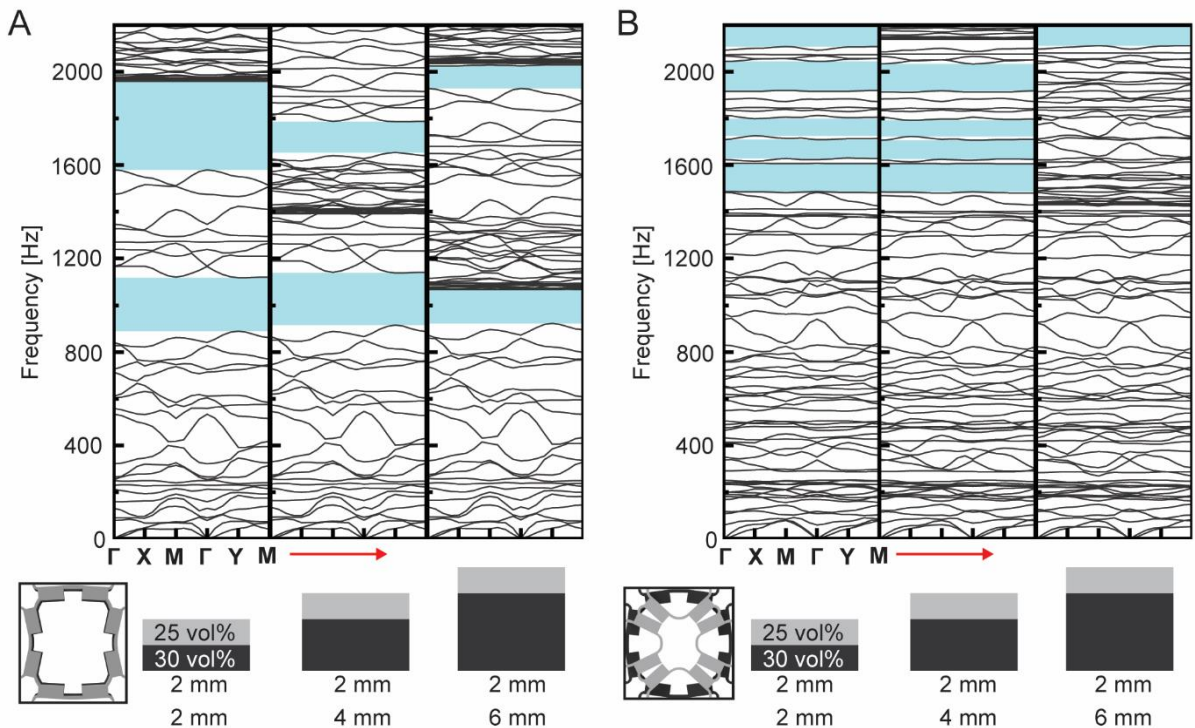

**Figure 5.** Effect of increasing bottom unit cell thickness on acoustic property. The top unit cell has a NdFeB particle loading of 25% and the bottom has 30%. The top unit cell thickness is kept constant while the bottom unit cell thickness increases. (A) Acoustic response of a unit cell in the initial mode with bottom unit cell thicknesses of 2 mm, 4 mm, and 6 mm. (B) Acoustic response of a unit cell in the top folded mode with bottom unit cell thicknesses of 2 mm, 4 mm, and 6 mm.

**2.4. Parametric Study of Flipped Bilayer**

In the previous section, the experimental bilayer was fabricated with an 8 × 8 bottom layer of 4 mm unit cells with 30 vol% NdFeB and a second 8 × 8 top layer of 2 mm unit cells with 25 vol% NdFeB. However, a flipped bilayer can be produced with unit cells with any thickness or particle loading. Thus, as shown in **Figure 5**, a parametric study of how unit cell thickness affects the acoustic response of a flipped bilayer assembly is conducted. The dispersion curves are calculated using a Bloch wave analysis on an RVE, identical to the simulation method in the previous section. See Figure S5 for more details. The top and bottom unit cells have 25 vol% and 30 vol% NdFeB, respectively. The top unit cell is kept at 2 mm while the bottom unit cell changes thickness from 2 mm to 4 mm, and then, to 6 mm. In addition, studies on unit cell thickness and the effect of particle loading on acoustic behavior can be found in Figure S9 and S10.

**Figure 5A** depicts the acoustic property modes when the flipped bilayer is in the initial mode. In each of the three presented acoustic property modes, there are two acoustic bandgaps, but the width and location vary greatly. When both top and bottom layers are 2 mm thick, the bandgaps are located at 889 Hz – 1118 Hz and 1580 Hz – 1958 Hz. At a bottom thickness of 4 mm, the bandgaps shift to 914 Hz – 1139 Hz and 1654 Hz – 1786 Hz. At a bottom thickness of 6 mm, the bandgaps are 923Hz – 1066 Hz and 1929 Hz – 2025 Hz. In all three cases, the low frequency acoustic bandgap is very consistent and only changes slightly in its width and location. However, the high frequency bandgap decreases in width as the bottom thickness increases. In addition, the lower limit of the bandgap increases in frequency. They can be utilized to design flipped bilayers with targeted bandgaps at certain width and location by using proper layer thickness.

In **Figure 5B**, the flipped bilayer unit cell is in the top folded mode. There are many acoustic bandgaps when the bottom unit cell thickness is set to 4 mm. The first combination, 2mm on bottom, has five acoustic bandgaps: 1488 Hz – 1606 Hz, 1630 Hz – 1709 Hz, 1726 Hz – 1799 Hz, 1921 Hz – 2043 Hz, and 2108 Hz – 2198 Hz. When the thickness of the bottom unit cell increases to 4 mm, the first four bandgaps display negligible change in location and width. However, the fifth bandgap disappears altogether. Therefore, there are only four bandgaps: 1486 Hz – 1605 Hz, 1628 Hz – 1706 Hz, 1723 Hz – 1791 Hz, and 1919 Hz - 2043Hz. When the bottom unit cell thickness is increased from 4 mm to 6 mm, all four acoustic bandgaps disappear and the fifth bandgap that dissappeared in the change from 2 mm to 4 mm reappears. This high frequency bandgap is from 2112 Hz to 2199 Hz, is close to the highest frequency bandgap when the bottom thickness equals 2 mm. Therefore, a metasurface with a thin bottom thickness will result in more acoustic bandgaps at a variety of frequency ranges in the folded state. As the thickness is increased, these bandgaps will disappear while the remaining bandgaps will remain relatively unchanged. This example demonstrate the great tunability of the flipped bilayer metasurface in acoustic properties.

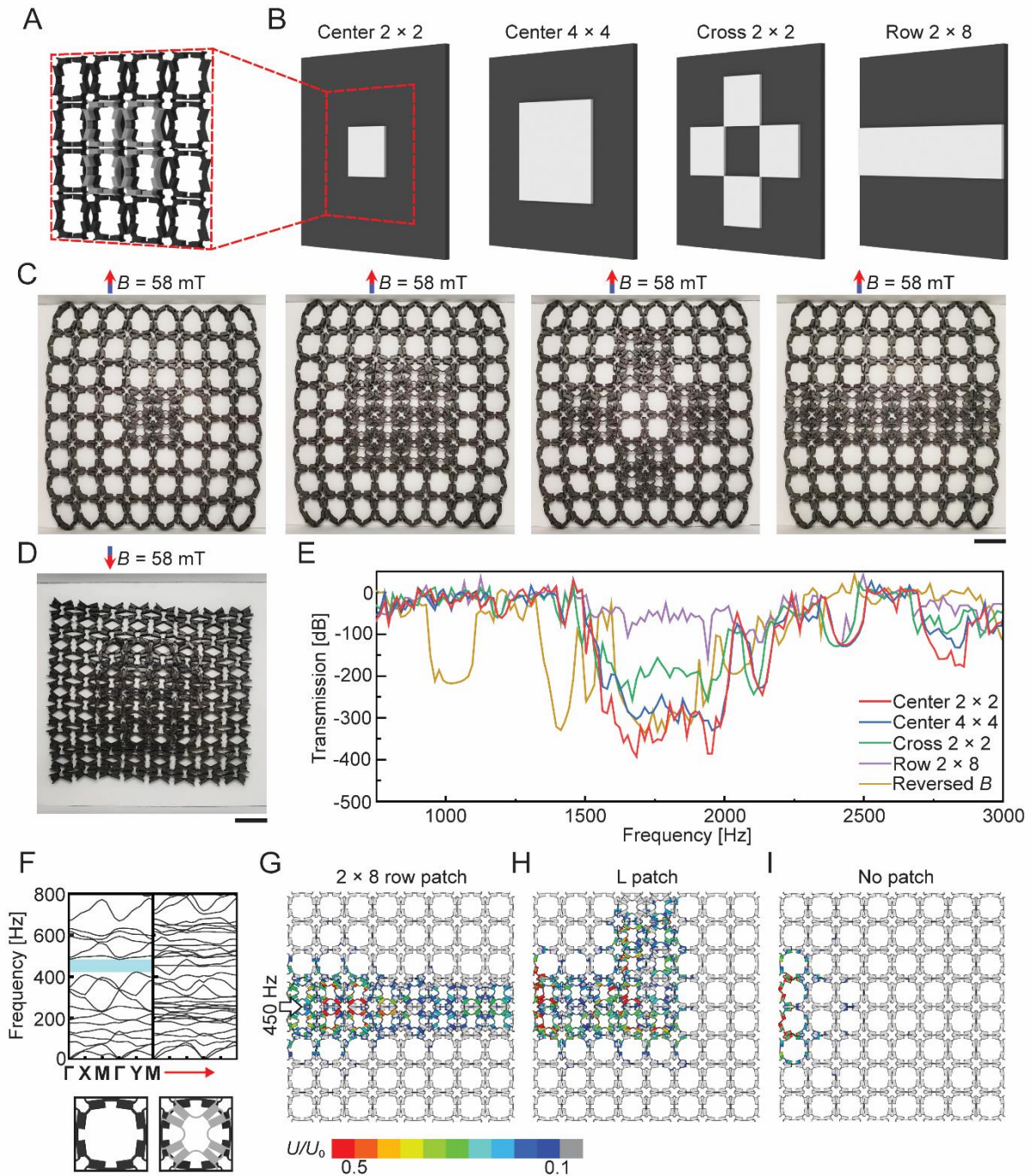

**Figure 6.** Flipped bilayer arranged in patch configurations and application as acoustic wave guide. (A) A patch configuration is where the top layer has a smaller number of rows or columns than the bottom layer. (B) Four patch configurations. (C) Magnetic actuation of the patch configurations. Scale bar: 20 mm. (D) Magnetic actuation for a center 4 × 4 patch on an 8 × 8 bottom layer under a downward magnetic field. Scale bar: 20 mm. (E) Transmission versus frequency curves of the four patch configurations when the field is 58 mT upwards and a single layer in the folded mode. (F) Acoustic property of the single layer in the initial mode and the flipped bilayer in the top folded mode. (G) Wave propagation on a 2 × 8 row patch. (H) Wave propagation on an "L" patch. (I) Wave propagation in a single layer in the initial mode, or no patch.

## 2.5. Rapidly Reconfigurable Flipped Patch

The flipped bilayer is produced using two different layers that have the same number of rows and columns. However, if the top layer has a smaller number of rows and columns, it can be rearranged arbitrarily on the bottom layer, like a patch. **Figure 6A** illustrates an example of a patch, where a top layer with two rows and columns is stacked on a bottom layer with four rows and columns. Simple schematics of the experimental flipped patch bilayers are illustrated in **Figure 6B**. Additionally, all configurations have an $8 \times 8$ bottom layer. The first configuration, called center $2 \times 2$, is a $2 \times 2$ patch at centered on top of the bottom layer. **Figure 6A**, is an enlarged view of the center $2 \times 2$ schematic. The second configuration is similar to the first, but has a $4 \times 4$ patch instead. Consequently, it is referred to as the center $4 \times 4$. The third patch configuration, or cross $2 \times 2$, utilizes four $2 \times 2$ patches and arranges them such that they form a cross with a $2 \times 2$ void in the center. Lastly, a $2 \times 8$ patch is placed along the width of the bottom layer. Although we investigate the four shown configurations, there is no limit to the number of rows and colums of the top array, or where it is placed. Therefore, the possible number of patch configurations, and correspondingly, property tuning ability, vastly increases.

**Figure 6C** shows the magnetic actuation of the studied patch configurations. The patches were fabricated using 2 mm thick unit cells and the bottom layers are 4 mm thick. Both patch and bottom have 30vol% NdFeB. An upward magnetic field of 58 mT acuates the patches and theycan be clearly distinguished by the contracted areas. **Figure 6D** shows an example of a patch configuration under a downward 58 mT magnetic field. Although the magnetic field forces the top patch to remain unfolded, the unit cells directly underneath the patch will fold, regardless of magnetic attraction between top and bottom. This is due to the size difference between the two layers. The bottom layer is much larger with total 64 unit cells, and only the center 16 are magnetically attracted to the top layer. The remaining 48 act as a single layer and exert large mechanical force to compact the unit cells at the center as well.

**Figure 6E** illustrates the transmission versus frequency curves of the four configurations and the center $4 \times 4$ under a downward 58 mT magnetic field. For the last patch configuration at this specific magnetic field, it is geometrically similar to the single layer in the compact mode (Figure 3C) and the acoustic property can be substituted as such. In all configurations except the $8 \times 2$ row, there are two noticeable transmission drops at the same frequencies. The only difference between each configuration is the depth of the transmission drop. The center $2 \times 2$ and the center $4 \times 4$ have similar deep transmission drops and the cross $2 \times 2$ has only about half of the former. In the case of the $2 \times 8$ row patch, there is no transmission drop. The 2 × 2 has the deepest transmission drop, followed by the center 4 × 4, and then the cross 2 × 2. At the frequencies where there are transmission drops, a single layer in the initial mode exhibits a bandgap, or deep transmission drop, and the flipped bilayer in the top folded mode does not. As more patches are added to the bottom layer, or single layer, the patch configuration will gradually behave similarly to the flipped bilayer. Therefore, the center 2 × 2 shows the deepest transmission drop. Although the other three patch configurations have the same number of unit cells in their patches, their acoustic behavior differs because of the patch shape. The cross 2 × 2 is discontinuous and the 2 × 8 row patch spans the entire length of the bottom layer.

**Figure 6F** illustrates the acoustic properties for the single layer deployed mode and the flipped bilayer with top folded mode. It shows a bandgap at 450 Hz for the single layer deployed mode, while the flipped bilayer with top folded shows a all-pass behavior. Therefore, by controlling the patch path, the metasurface can serve as a waveguide to effectively regulate the wave propagation direction. To demonstrate this concept, we simulated the wave propagation for different patch paths at 450 Hz using FEA (See supporting information for details on FEA simulation). **Figure 6G** shows the wave propagation of a 2 × 8 row patch configuration, illustrating the propagation of the acoustic wave only along a straight line, following the patch path. **Figure 6H** shows the wave propagation that is bent at a right angle, achieved by an L-shaped patch. Conversely, **Figure 6I** depicts that acoustic wave fails to propagate in the single layer metasurface at its deployed mode. These examples demonstrate the viability of a patch arrangement, to expand the design space and as wave guides.

## 3. Conclusion

We report magneto-mechanical bilayer metasurfaces that achieve excellent shape and property tunability through a novel strategy, where the unit cells are arranged in double layer patterns. This new design method allows for the metasurfaces to adjust their area density with negligible change to its overall dimensions. We presented three bilayer strategies: an offset bilayer, flipped bilayer, and patch configurations. When actuated with an applied magnetic field, the offset bilayer contracts inwards and shows improved area preservation and property tunability than of a single layer metasurface. The flipped bilayer utilizes separate actuation of its two layers to achieve increased shape transformations, property tunability, and exceptional area preservation. Lastly, the patch configuration greatly expands the design space and can be utilized as a reconfigurable wave guide. This novel strategy in utilizing two layers to simultaneously achieve shape deformation and area preservation offers immense potential for new metasurfaces with better property controllability. The design of the magneto-mechanical metasurface can be further guided by machine learning strategies for optimized structural design[38] and actuation programming to achieved desired property tunability.

## 4. Experimental Section

*Unit Cell Preparation*: In this study, three different unit cells were used to create the metasurfaces. In all three cases, Dragon Skin 20 (Smooth-On Inc., Macungie, PA, USA) and NdFeB particles (average particle size of 100 µm, Magnequench, Singapore) were combined, degassed, and injected into polyvinyl alcohol (PVA) molds 3D printed with an Ultimaker S5 (Ultimaker, Netherlands). The filled molds were covered with a glass slide and cured at 80 ºC for 1 hour. PVA is flexible at high temperatures, allowing the cured unit cells to be removed easily. The first unit cell was 4 mm thick and had a particle loading of 30 vol%. The other two cells were both 2 mm thick with 25 vol% and 30 vol% NdFeB. To achieve the magnetization shown in **Figure 1A**, the unit cells were deformed into a "diamond" shape and held in place with a 3D printed polylactic acid (PLA) fixture. This mold was placed in a home-made coil and magnetized under a 1.5 T impulse field. More details on the unit cell fabrication process and dimensions can be found in the supporting information.

*Metasurface Array Fabrication*: After magnetization, the unit cells were placed into a 3D printed PLA fixture that allowed the hinges of each unit cell to touch the other. The hinges were connected using Sil-poxy adhesive (Smooth-On Inc., Macungie, PA, USA). The unit cells were removed from the fixture after the adhesive was allowed to cure at room temperature for twelve minutes.

*Acoustic Simulation*: The acoustic properties of each deformation mode in all metasurface systems were predicted using commercial software ABAQUS (Dassault Systèmes SE, Vélizy-Villacoublay, France) and COMSOL (COMSOL Inc., Stockholm, Sweden). A steady state dynamic analysis in ABAQUS was performed on metasurface systems with non-periodic structures such as the offset bilayer and patch bilayer. However, for the flipped bilayer, where the structure is periodic throughout, a Bloch wave analysis was used to predict the acoustic properties using COMSOL.

*Experimental Setup*: The metasurfaces were placed between a pair of Helmholtz coils spaced 150 mm apart. The coils were connected to a 4700 µF capacitor which in turn, is connected to a 120 V wall outlet. The magnetic field between the coils was controlled by adjusting the current delivered to the coils using a MATLAB script. Further information on the experimental setup can be found in the supporting information.

*Material Characterization*: Like the fabrication process for the unit cells, Dragon Skin 20 and NdFeB particles were mixed into a homogenous mixture in specific volume percentages and degassed. The mixture was then poured into acrylic molds for tensile test specimens. The dimensions of the test specimens can be found in the supplementary information. The mixture was then cured in an oven at 80 ºC for 1 hour and the molds were removed from the test specimens. An Instron 3344 (Instron Corp., Norwood, MA, USA) was used to characterize the displacement vs. force data, and the data was fitted to a neo-Hookean model to estimate the Young's modulus. This process was done three times for tensile test specimens with particle loadings of 20 vol%, 25 vol%, and 30 vol%.

*Area Density Calculation*: Actuation videos were recorded of the fabricated metasurfaces placed inside the magnetic coils and actuated at specific magnetic fields. Image analysis of the metasurfaces were conducted by cropping image snapshots of the videos and calculating the total number of pixels in $A_{meta}$, $A_{initial}$, and $A_{deform}$, and then calculating the global and local area densities.

**Acknowledgements**
The authors acknowledge the support from NSF Career Award CMMI-2145601 and NSF Award CMMI-2142789.